\documentstyle[floats,aps]{revtex}
\begin{document}
\input epsf
\draft

\twocolumn[\hsize\textwidth\columnwidth\hsize\csname  
@twocolumnfalse\endcsname
\title{Finite Size Scaling Behavior of 
 Dissipative Tunnel Junctions}

\author{S. Drewes and S.R. Renn}
\address{
Department of Physics, University of California at San Diego,
La Jolla, CA 92093}

\date{\today}

\maketitle

\begin{abstract}
We present the results of a series of Quantum Monte Carlo calculations of the 
temperature dependent conductance of dissipative non-superconducting tunnel
junctions.  Finite size scaling methods are used to demonstrate the absence of 
coherent transport in Ohmic tunnel junctions.  However, a quantum phase 
transition between coherent and incoherent behavior is found in sub-Ohmic tunnel
junctions.  The critical conductance is found to be temperature independent
and the critical exponents, $\nu$ and $\eta$, are found to be in agreement with
renormalization group results by Kosterlitz.
\end{abstract}

\vskip2pc]

\narrowtext

In spite of the significant effort to understand the effect of ohmic
dissipation on normal (non-superconducting) tunnel junctions, there remains a
conflict between analytical studies\cite{Kosterlitz,Fisher,largeN} and quantum
Monte Carlo simulations\cite{Scalia,Zwerger} of these systems.  In particular,
analytical treatments such as renormalization group\cite{Kosterlitz}, large N 
approximation\cite{largeN}, and variational methods\cite{Strohm} generally 
conclude that no quantum phase transition occurs in tunnel junctions subjected 
to a purely Ohmic environment.  In contrast, it has been argued\cite{Scalia} 
that quantum Monte Carlo studies of Ohmic tunnel junctions exhibit a quantum 
phase transition between an ordered phase which exhibits coherent tunneling and
a disordered phase which does not.

If such a phase transition existed, it would be a remarkable example of a
quantum phase transition driven by dissipation.  Moreover, the existence of
such a transition would be of considerable  importance to the electronic and 
transport properties of a variety of
systems including double quantum dots and metallic granular composites.
For these reasons, we believe that it is essential to resolve this discrepancy.
Towards this end, we  will reexamine the simulation data using 
using finite size scaling methods.  This approach will allow us to
identify quantum phase transitions which may occur in Ohmic and sub-Ohmic
tunnel junctions.

Our discussion is based on the generalized
Ben Jacob Mottola Schon (BMS) model\cite{BenJacob} which describes the
effects of dissipation and Coulomb blockade phenonenon on non-superconducting
tunnel junctions.  The BMS model is defined by the effective action
\begin{equation}
S[\phi]= \int d\tau\ d\tau' \alpha(\tau-\tau')[1-\cos(\phi(\tau)-\phi(\tau'))]
\end{equation}
where the phase $\phi(\tau)$ is defined in terms of the time-dependent voltage 
across the tunnel junction, $V(\tau)$, such that $V(\tau)=\frac{e}{\hbar} 
\dot{\phi}(\tau)$.
The generalized BMS model 
is a long range XY model with spin-spin interactions of the form $\alpha(\tau)=
\alpha_0\tau_Q^{-\epsilon}\left( \frac{\pi k_B T}{\sin(\pi k_B T 
\tau)}\right)^{2-\epsilon}$. 
$\tau_Q$ is defined in terms of the charging energy $E_Q=e^2/2C$ by  
$\tau_Q=\hbar/E_Q$. Charging effects are introduced to the model by working on
a lattice such that $\phi(\tau)$ is defined for $\tau=j \tau_Q$ where 
$j=1 \dots N$ and where $N=\hbar \beta/\tau_Q=E_Q/k_BT$ is the total number of 
timeslices.
$\epsilon$ is a parameter, discussed below, which characterises dissipative 
bath coupled to the tunnel junction.
When $\epsilon=0$, the coupling parameter, $\alpha_0$, is related to  $R_T$, 
the $T\rightarrow \infty$ limit of the resistance, by the expression 
$\alpha_0=\hbar/(2\pi e^2 R_T)$.  More generally $\alpha_0$ is proportional to 
$|t|^2 N_LN_R$ i.e.  the squared tunneling matrix element multiplied by the 
density of states of the left and right electrodes.

When one derives this action from a more microscopic
model, the  long range interaction,  $\alpha(\tau)$,
is generated by  intergrating out the modes associated
with a  dissipative bath.  These modes might, for instance, represent, 
particle-hole excitations
occuring in the tunnel junction electrodes. In the absence of manybody effects,
the particle-hole excitations give rise to Ohmic dissipation ($\epsilon=0)$.  
However, in some situations, Fermi edge singularity effects such as the 
orthogonality catastrophe and the exciton effect\cite{mahan,Noziers}  
may become important.\cite{Drewes,Ueda}. In such cases, the dissipative bath is
not Ohmic i.e. $\epsilon \ne 0$ and the tunnel junction may exhibit a variety 
of interesting effects
including  non-linear
$I(V)$ characteristics.

  As discussed by Drewes et al\cite{Drewes}, the
renormalization group theory  predicts a quantum phase transition for sub-Ohmic
tunnel junctions (i.e. $\epsilon>0$). This transition occurs at  
$\alpha_0=\alpha_c(\epsilon)$ where $\alpha_c(\epsilon)\rightarrow \infty$ as 
$\epsilon \rightarrow 0$. This result appears to be in 
conflict with the Monte Carlo simulations
of Ohmic ($\epsilon=0$) tunnel junctions by Scalia et al\cite{Scalia}. The 
latter study interpreted the observed $\alpha_0$ dependence
of the correlation
function $g(\tau)=<\exp i\phi(\tau) \exp-i\phi(0)>$ 
as evidence for a quantum
phase transition   at $\alpha_0^{-1}\approx 0.7$.

In this paper, we will reconsider the simulation results using
finite size scaling methods. Using these methods we have obtained the following
results: (1.) No quantum phase transition occurs in Ohmic
tunnel junctions with $\alpha_0<10.0$; (2.)  The sub-Ohmic tunnel junctions 
exhibit
a quantum phase transition characterized by a temperature independent critical 
conductance of order $e^2/h$;  (3.)  Values for the 
$\eta$ and $\nu$ critical exponents are found to be 
consistent with the analytical results of Kosterlitz\cite{Kosterlitz} and 
Nickel et al\cite{Fisher}.

\vskip 0.15truein

\noindent {\it A Quantum Phase Transition:}  We begin our discussion, by 
considering the tunnel junction conductance which can be written 
as\cite{Drewes,Simanek}
\begin{equation}
G=\frac{2\pi \alpha_0}{\hbar \beta R_Q}\int_0^{\hbar \beta}
\gamma_{\epsilon}(\tau)\langle \cos(\phi(\tau)-\phi(0))\rangle \ d\tau
\label{GDC}\end{equation}
where $\gamma_{\epsilon}(\tau)=[\pi(k_BT/E_Q)/\sin(\pi k_B T \tau 
\hbar)]^{-\epsilon}$.  We evaluate $G$ using
$g(\tau)=\langle \cos(\phi(\tau)-\phi(\tau'))\rangle$
obtained from a series of simulations.  The simulations involved the standard 
Metropolis Rosenbluth Teller Teller algorithm of the long range XY model on a 
1-D lattice of $N=\hbar \beta/\tau_Q$ spins and periodic boundary conditions.
Our simulations  included $10^5$ cycle runs for the
10, 20, and 32 timeslice systems and $2\times 10^5$ cycle runs for the 64 and
128 timeslice systems.

\begin{figure}[!h]
\centering
\leavevmode
\epsfxsize=9cm
\epsfysize=9cm
\epsfbox[12 12 576 480] {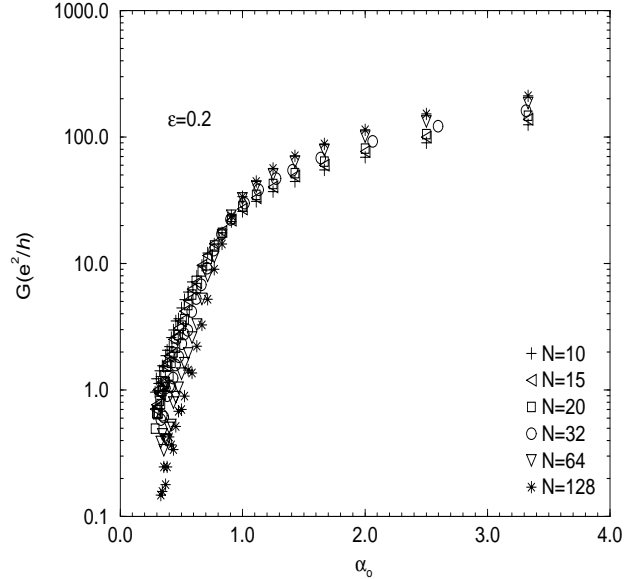}
\caption[]
{\label{Gpttwo} Conductance for $\epsilon=0.2$. Notice that the conductance 
curves cross at $G_c\approx 21e^2/h$. The crossing separates a phase where the 
conductance exhibits sub-Ohmic from a phase where it exhibits insulating 
behavior.}
\end{figure}

\begin{figure}[!h]
\centering
\leavevmode
\epsfxsize=9cm
\epsfysize=9cm
\epsfbox[12 12 576 480] {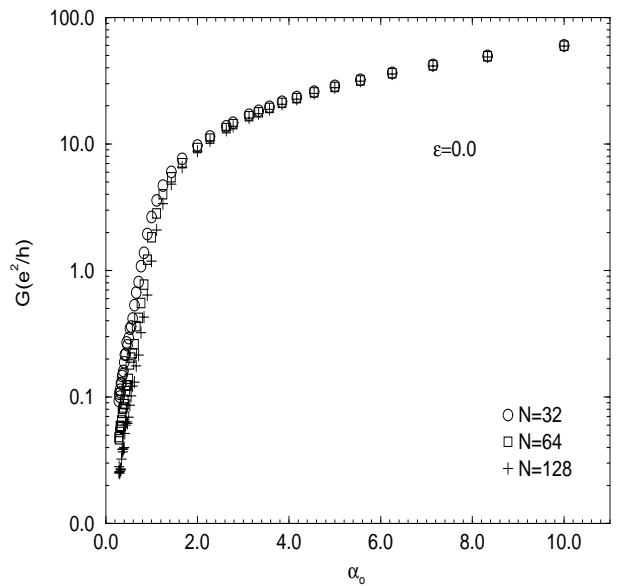}
\caption[]
{\label{Gzero} Conductance for $\epsilon=0.0$.}
\end{figure}

The results for $\epsilon=0.2$ are presented in
fig. \ref{Gpttwo} and the results for $\epsilon=0$ are presented in fig. 
\ref{Gzero}.  Interestingly, in fig. \ref{Gpttwo} the $N$ dependence of
the conductance curves reverses at the point of intersection at 
$\alpha_c\approx 0.9$.  For $\alpha_0>\alpha_c$ one observes metallic behavior 
i.e. a regime where $G$ increases  as $k_BT=E_Q/N \rightarrow 0$. Conversely,
for $\alpha_0<\alpha_c$, $G$ decreases as $T\rightarrow 0$.  Hence, we may 
identify the crossing point as a quantum
phase transition between two phases.  Hereafter, these two
phases will be referred to as sub-Ohmic ( $\alpha_0>\alpha_c$) 
or insulating ($\alpha_0<\alpha_c$).
The sub-Ohmic phase exhibits a conductance which increases
with increasing $N \propto 1/k_BT$ whereas the insulating
exhibits a  conductance which  decreases with increasing N.

\begin{figure}[!h]
\centering
\leavevmode
\epsfxsize=9cm
\epsfysize=9cm
\epsfbox[18 18 552 482] {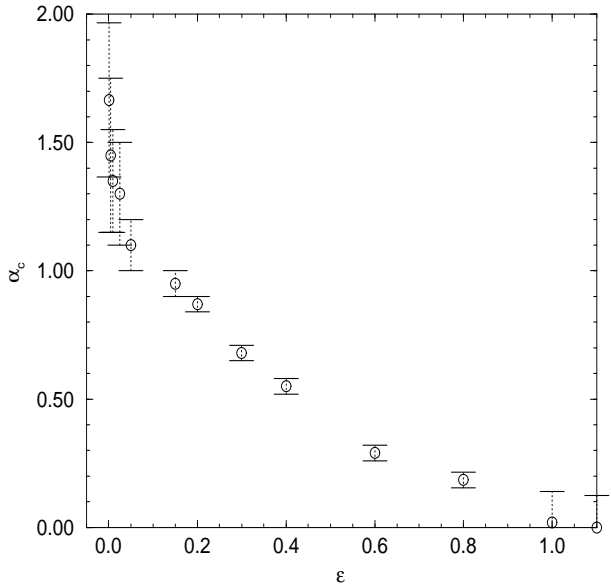}
\caption[]
{Critical value of $\alpha_0$ vs $\epsilon$ as determined by the crossing
of conductance curves.}\label{alphacrit}
\end{figure}

Remarkably the conductance curves in Fig. \ref{Gpttwo}  don't simply exhibit
a set of crossings. Instead all six curves cross at a single
point.  This observation suggests that the critical conductance is 
temperature independent.  The existence of a temperature independent
critical conductance can
be understood using the finite size scaling ansatz 
\begin{equation}
<e^{i \phi(\tau)}e^{-i\phi(0)}> = 
\left( \frac{\tau_Q}{\tau}\right)^{d-2+\eta}K_{\pm}
\left(\frac{L_{\tau}}{\xi_{\tau}}, \frac{\tau}{\xi_{\tau}}\right)
\label{correl}\end{equation}
where $d=1$ is the dimensionality of the model, $\tau_Q=\hbar/E_Q$ is the size 
of the time slices, and where $\xi_{\tau}$ is the correlation time.
Eq. \ref{correl} together with eq. \ref{GDC} determine the scaling behavior of 
the conductance:
\begin{equation}
G=\frac{e^2}{h}\left( \frac{k_BT}{\Delta}\right)^{\eta-1-\epsilon} 
F_{\pm}\left(\frac{k_BT}{\Delta}\right)
\label{Gscaling}\end{equation}
where $\Delta\equiv \hbar/\xi_{\tau}$ and where $F_+(x)$ and $F_{-}(x)$ are 
the $\alpha_0<\alpha_c$ and $\alpha_0>\alpha_c$ branches of some universal 
scaling function.  Now  Fisher, Ma and Nickel
\cite{Fisher} have shown that the long range spin-spin interactions cause 
$\eta=1+\epsilon$.  Because of this, eq. \ref{Gscaling}  implies a temperature 
independent 
critical conductance, $G_c(\epsilon)$, of order $e^2/h$.
This result  has been confirmed using 
a large N approximation of the model\cite{largeN}.  According to that treatment
\begin{equation}
G_c(\epsilon)=2\pi (1-\epsilon) {\rm ctn} \left( \frac{\epsilon \pi}{2} 
\right) \frac{e^2}{h}
\label{largeN}\end{equation}
This result shows that $G_c\rightarrow 0$ as $\epsilon\rightarrow 1$ which 
indicates that the disordered phase is absent at $\epsilon=1$.
Similarly eq. \ref{largeN} shows that  $G_c\rightarrow \infty$ as $\epsilon 
\rightarrow 0$.  This suggests that the ordered phase is absent at 
$\epsilon = 0$. 
These interpretations
are supported by the observed $\epsilon$ dependence of $\alpha_c(\epsilon)$ 
given in  fig. \ref{alphacrit}.

\begin{figure}[!h]
\centering
\leavevmode
\epsfxsize=9cm
\epsfysize=9cm
\epsfbox[12 12 576 480] {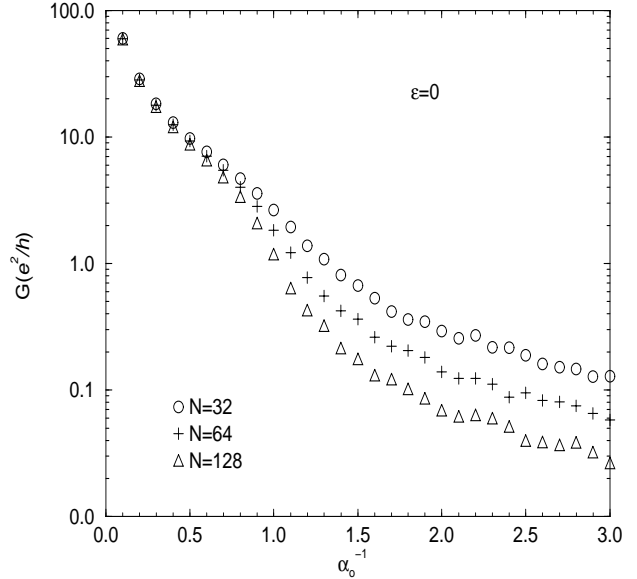}
\caption[]
{\label{grainb}Same data as in  previous figure, except G is plotted vs 
$1/\alpha_0$.  Observe that the conductance is monotonically increasing 
function of temperature , $k_BT \propto E_Q/N$. }
\label{Galpinv}\end{figure}

As was mentioned in the introduction, the absence of an ordered phase in the
$\epsilon \rightarrow 0$ limit is not only a prediction of the large $N$ 
approximation but also is predicted by Kosterlitz' renormalization group 
treatment
of the long-range XY model.\cite{Kosterlitz}  Additional evidence 
may be found in figs. \ref{Gzero} and \ref{Galpinv}. 
For  $10>\alpha_0>0.1$, $G$ decreases with decreasing
temperature, $k_BT=E_Q/N$.  This strongly suggests that no metallic or coherent
phase is occurs in this range of $\alpha_0$.  In addition, these figures
do not exhibit any crossings which might indicate a
phase transition.

\vskip 0.15truein

\begin{figure}[h]
\centering
\leavevmode
\epsfxsize=9cm
\epsfbox[18 18 552 482] {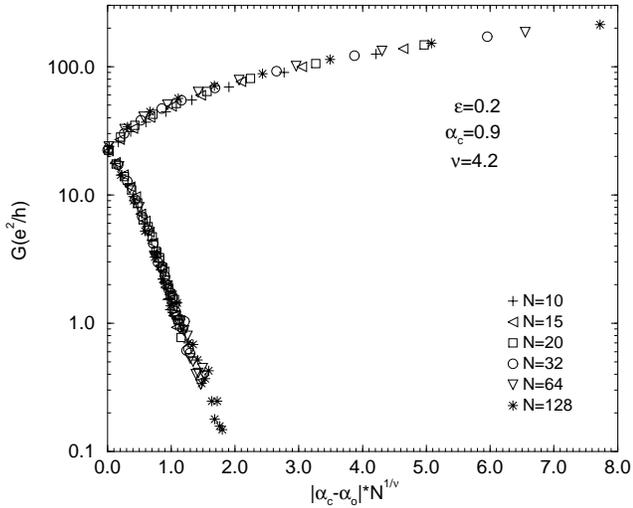}
\caption[]
{\label{cond.min.2}The conductance at $\epsilon=0.2$.  Two branches occur. 
The upper branch is metallic (increasing with N) and the lower branch is 
insulating
(decreasing with N).  Data collapse validates scaling form in eq. 
\ref{Gscaling}.}
\label{scalepttwo}
\end{figure}

\noindent {\it Finite Size Scaling Behavior:}
  We have seen that the existence of the temperature independent conductance 
provides
indirect evidence of the finite size scaling behavior proposed in eq. 
\ref{Gscaling}.
To obtain  more direct evidence for finite size scaling, one must be able to
to collapse the conductance data using
\begin{equation}
G=\frac{e^2}{h}\tilde{F}_{\pm}\left( N^{1/\nu}|\alpha_0-\alpha_c|\right)
\label{scalecc}\end{equation}
This is done in fig. \ref{scalepttwo}. In that figure, we used 
the value of $\alpha_c$ from fig. \ref{Gpttwo} and adjusted $\nu$ to minimize 
the scatter. In this manner we obtained data collapse with $\nu=4.2\pm 0.6$ for
$\epsilon=0.2$
This may be compared with the one loop renormalization group result that 
$\nu=1/\epsilon[1+O(\epsilon)]$.

\begin{figure}[!h]
\centering
\leavevmode
\epsfxsize=9cm
\epsfysize=9cm
\epsfbox[18 18 552 482] {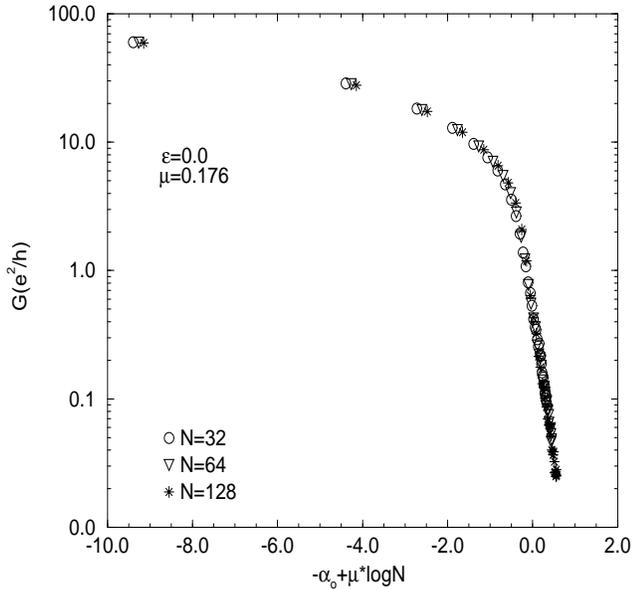}
\caption[]
{\label{grain} Conductance for $\epsilon=0.0$. Note: The conductance is 
monotonically increasing function of temperature,  $k_BT =E_Q/N$.
This implies insulating behavior at low temperatures.}
\label{scalezero}\end{figure}

Next consider the $\epsilon=0$ limit.  In this limit,
the scaling behavior indicated in eqn. \ref{scalecc} reduces to
\begin{equation}
G=\frac{e^2}{\hbar}L(\alpha_0 -\mu \ln N)
\end{equation}
where $\mu=\lim_{\epsilon \rightarrow 0} \alpha_c/\nu$ is assumed to be finite.
The resulting data collapse occurs is displayed  in fig. \ref{scalezero}.
A comparison between figs. \ref{scalepttwo}  and \ref{scalezero} indicates that
only the monodecreasing branch of the scaling function is present in the  
$\epsilon=0$ data.  This means that for all $\alpha_0$,
the conductance of Ohmic tunnel junctions decreases with increasing 
$N\propto 1/k_BT$.  Again we conclude
that, as predicted by Kosterlitz's renormalization group treatment,
a coherent (ordered) phase is
does not occur in Ohmic tunnel junctions.

\noindent {\it Summary:} We have performed a finite size scaling analysis of 
the conductance of disspative Ohmic and sub-Ohmic tunnel junctions.  Consistent 
with the predictions of renormalization group and large N theories, the 
sub-Ohmic tunnel junction exhibits a quantum phase transition between 
sub-Ohmic and insulating phases.  The results for Ohmic tunnel junctions are 
also consistent with
the RNG predictions:  In particular, little evidence for a metallic phase is 
found in Ohmic tunnel junctions for $\alpha_0<10$. However,
the conductance does exhibit finite size scaling consistent with a quantum 
phase transition which, according to renormalization group theory and large N 
approximation, occurs at $\alpha_c^{-1}=0$ in the Ohmic ($\epsilon=0$) limit.

\noindent {Acknowledgements:} The authors would like to acknowledge support from
NSF Grant No. DMR 91-13631 (S.R.),  the Hellman foundation(S.R.), 
the Alfred P. Sloan Foundation (S.R.). We would like to acknowledge
useful conversations with F. Guinea and D.P. Arovas.


\begin{references}

\bibitem{Kosterlitz} J.M. Kosterlitz,  Phys. Rev. Lett. {\bf 37}, 1577 (1977).
\bibitem{Fisher}M.E. Fisher, Shang-keng Ma, B.G. Nickel, Phys. Rev. Lett.
{\bf 29}, 917 (1972).
\bibitem{largeN}S.R. Renn, UCSD preprint 1997.
\bibitem{Scalia}V. Scalia, G. Falci, R. Fazio, G. Giaquinta, Z. Phys. B.
{\bf 85}, 427-433(1991).
\bibitem{Zwerger} W. Hofstetter and W. Zwerger, Phys. Rev. Lett.
{\bf 78}, 3737 (1997).
\bibitem{Strohm}T. Strohm and F. Guinea, Nucl. Phys. B {\bf 487}, 795 (1997).

\bibitem{BenJacob}E. Ben-Jacob, E. Mottola, and G. Sch$\ddot{o}$n, Phys. Rev.
Lett. {\bf 51}, 2064 (1983).


\bibitem{Ueda}M. Ueda and F. Guinea, Z. Phys. B {\bf 85}, 413 (1991);
 M. Ueda and S. Kurihara in Macroscopic quantum phenomena,
T.D. Clark, H. Prance, R.J. Prance, T.P. Spiller (eds.), p.143, Singapore,
World Scientific (1990);T. Strohm and F. Guinea, Nucl. Phys. B {\bf 487}, 795 
(1997).

\bibitem{mahan} G.D. Mahan, Phys. Rev. {\bf 153}, 882 (1967); G.D. Mahan,
Phys. Rev. {\bf 163}, 612 (1967).

\bibitem{Noziers}P. Nozi\'eres and C. T. De Dominicis, Phys. Rev. {\bf 178},
1097 (1969).






\bibitem{Drewes}S. Drewes, S. R. Renn, F. Guinea, UCSD preprint.



\bibitem{Simanek} This result generalizes an expression obtained for the 
$\epsilon=0$ limit by  R. Brown and E. Simanek, Phys. Rev. B {\bf 34}, 2957 
(1986).










\end{references}
\end{document}